\documentclass[aps,amsfonts,amsmath,prd,preprint,nofootinbib]{revtex4}

\newcommand{\beq}{\begin{equation}}
\newcommand{\eeq}{\end{equation}}

\usepackage{epsfig}
\usepackage{graphicx,color}

\begin{document}

\title{{\Large Skyrme Branes}}

\author{{\large Jose J. Blanco-Pillado}$^{a,}$}
\email{jose@cosmos.phy.tufts.edu}
\author{{\large Handhika S. Ramadhan}$^{a,}$}
\email{Handhika.Ramadhan@tufts.edu}
\author{{\large Noriko Shiiki}$^{b,}$}
\email{norikoshiiki@mail.goo.ne.jp}

\affiliation{$^a${\em Institute of Cosmology,
Department of Physics and Astronomy\\ Tufts University, Medford, MA 02155}}
\affiliation{$^b${\em Graduate School of Art and Science,
University of Tokyo, Komaba 3-8-1, Tokyo 153-8902, Japan}}

\begin{abstract}
We obtain static selfgravitating solitonic 3-brane solutions in the 
Einstein-Skyrme model in $7D$. These solitons correspond to a smooth version of the
previously discussed cosmic p-brane solutions. We show how the energy
momentum tensor of the Skyrme field is able to smooth out the
singularities found in the thin wall approximation and falls fast
enough with the distance from the core of the object so that
asymptotically approaches the flat cosmic p-brane metric.

\end{abstract}

\maketitle

\section{Introduction}
Models with extra dimensions have become intensively studied in the
last few years, not only in string theory but in many other extensions
of the standard model. Branes have played a major role on many of
these models and have therefore been studied from several different
perspectives \cite{BRANES}. Some of these branes have a description in
terms of solitons in the low energy theories as either domain walls \cite{domainwalls},
strings \cite{strings}, monopoles \cite{monopoles}, or even higher
codimension objects \cite{generalcodimension}. One of the
earliest models of particle physics to incorporate solitons to its
spectrum, the Skyrme model \cite{Skyrme}, has so far not been investigated in connection
to the braneworld scenarios. The purpose of this letter is to study
braneworld models based on the higher dimensional generalization of the
Einstein-Skyrme model.

One of the most interesting properties of the Skyrme model that makes
it different from all the other topological defects mentioned above, is
the presence of higher order corrections to the kinetic term. Recently
there has been several studies on topological defects with
non-canonical kinetic terms \cite{k-defects}. Most of those models have
been focusing on the differences introduced in the soliton solutions
by the new term in the lagrangian. On the other hand, our model does
not have a potential term in contrast to all the other cases studied
in the literature of braneworld scenarios. Furthermore, in the absence of the
Skyrme term, the theory does not present any smooth stable
configuration even though there is a topological charge that one can 
define. It is only because of this new term that one is able to have 
a finite size stable configuration avoiding in this way the
straighforward extension of Derrick's theorem \cite{Derrick}. It is 
then clear that the higher order terms in our Lagrangian are crucial 
for the solutions presented here and are not just a small correction to the model.

On the other hand, the particular structure of the theory that we study
here gives rise to solitonic solutions that are very much localized
in the transverse directions. This property makes them good candicates to
describe a smooth version of extended solutions previously studied
only in the their thin wall limit. In fact,
we will show that our thick brane solutions in the Einstein-Skyrme
model asymptotically match the higher dimensional vacuum solutions of
the pure Einstein's equations previously found in \cite{RG}.

\section{The Einstein-Skyrme model in 7D}

The action for our model is given by,
\begin{equation}
S_{ES}=\int d^{7}X\sqrt{-g}\left[\frac{1}{2\kappa^{2}}R+\cal{L_{S}}\right],
\end{equation}
where $R$ is the Ricci scalar, $\kappa^2=1/M_7^5$, with $M_7$ denoting
the 7-dimensional Planck mass, and $\cal{L_{S}}$ is the Skyrme
Lagrangian density,
\begin{equation}
{\cal{L_{S}}}=\frac{F_{0}^{2}}{4}Tr(L_{A}L^{A})+\frac{1}{32e^{2}}Tr([L_{A},L_{B}][L^{A},L^{B}]),
\end{equation}
where $F_{0}$ and $e$ are two free parameters of the model with units
of $[M]^{5/2}$ and $[M]^{-3/2}$ and
\begin{equation}
L_{A}\equiv U^{\dag}\partial_{A}U,
\end{equation}
is the left chiral current and $U\in SU(2).$\footnote{We use the
  following notation. The upper case latin indices $A,B$ run over
$0,..,6$ and the greek indices $\mu, \nu = 0,...,3$ denote the four
dimensional spacetime coordinates. We use the the mostly positive
signature and the Riemann tensor conventions of the form, $R^A_{BCD} = \partial_C
\Gamma^A_{BD} - \partial_D \Gamma^A_{BC} + ...$.}

We are interested in finding the smooth solution for a 3-brane
that is spherically symmetric along the transverse directions
in the bulk. We will also restrict ourselves to the four dimensional
flat brane solutions. Taking these constraints into account we
can now write the most general metric of this form in the
{\it isotropic gauge} as,

\begin{equation}
\label{eq:metric}
ds^{2}=B^{2}(r)\eta_{\mu\nu}dy^{\mu}dy^{\nu}+C^{2}(r)\left(dr^{2} + r^{2}d\Omega^{2}\right),
\end{equation}
where $\eta_{\mu\nu}=diag(-1,1,1,1)$.

We also impose the hedgehog ansatz for the chiral field, which
is the natural spherically symmetric ansatz for Skyrme model; that is, we assume
the following form for $U(r)$,
\begin{equation}
U(r)=\cos f(r)+i \left(\frac{r^j}{r}\right) \tau^j \sin f(r),
\end{equation}
where $f(r)$ is the profile function to be solved for and $\tau^j$ with
$j=1,2,3$, are the Pauli matrices.

Within this ansatz, the Lagrangian density then becomes,
\begin{eqnarray*}
{\cal{L_{S}}}&=&\frac{-F_{0}^{2}}{2}\left[\frac{1}{C^{2}(r)}\left(\frac{df}{dr}\right)^{2}
\left(1+\frac{2\sin^{2}f}{e^{2}F_{0}^{2}C^{2}(r)r^{2}}\right)
+\frac{\sin^{2}f}{C^{2}(r)r^{2}}\left(2+\frac{\sin^{2}f}{e^{2}F_{0}^{2}C^{2}(r)r^{2}}\right)\right].
\end{eqnarray*}
It is convenient at this point to rescale the radial coordinate
$eF_{0}r\rightarrow x$ and define,
\begin{eqnarray*}
\label{eq:uandv}
u&\equiv& \frac{1}{C^{2}(x)}\left(1+\frac{2\sin^{2}f}{C^{2}(x)x^{2}}\right)~,\nonumber\\
\\
v&\equiv&\frac{\sin^{2}f}{C^{2}(x)x^{2}}\left(2+\frac{\sin^{2}f}{C^{2}(x)x^{2}}\right)~,
\end{eqnarray*}
to obtain
\begin{equation}
{\cal{L_{S}}}=\frac{-e^{2}F_{0}^{4}}{2}\left(u f'^{2}+v\right).
\end{equation}
Having simplified the Lagrangian, the action for the Skyrme field becomes,
\begin{eqnarray}
S_{S}&=&\int d^{7}X\sqrt{-g}\cal{L_{S}}\nonumber\\
&=&\frac{-2\pi F_{0}}{e}\int(uf'^{2}+v)B^{4}(x)C^{3}(x)x^{2}dx d^{4}y.
\end{eqnarray}
It is now straightforward to obtain from this action the equation of
motion for the field $f(x)$ in the static case which is given by,
\begin{equation}
\label{eq:f}
f''(x)=\frac{1}{2u}\left(u_{f}f'^{2}(x)+v_{f}\right)-
\left[4\frac{B'(x)}{B(x)}+3\frac{C'(x)}{C(x)}+\frac{u'}{u}+\frac{2}{x}\right]f'(x),
\end{equation}
where
\begin{eqnarray}
u_{f}&\equiv&\frac{\delta u(x)}{\delta f(x)},\nonumber\\\\
v_{f}&\equiv&\frac{\delta v(x)}{\delta f(x)},
\end{eqnarray}
and the primes denote the derivatives with respect to $x$. On the other hand,
varying the Skyrme action with respect to the metric tensor $g_{AB}$ yields
\begin{equation}
T_{AB} = -\frac{2}{\sqrt{-g}} \frac {\delta S_S} {\delta g^{AB}}~,
\end{equation}
where $T_{AB}$ is the energy-momentum tensor for the scalar field given by
\begin{eqnarray*}
\label{eq:emtensors}
T_{AB}&=&g_{AB}{\cal{L_{S}}}-\frac{F_{0}^{2}}{2} Tr(L_{A}L_{B})-\frac{1}{8e^{2}}~g^{MN}Tr([L_{A},L_{M}][L_{B},L_{N}]).
\end{eqnarray*}
In our ansatz it becomes,

\begin{eqnarray}
T^{\mu}_{\nu}      &=& - \frac{e^{2}F_{0}^{4}}{2}\left(u f'^{2}+v\right)\delta^{\mu}_{\nu}\nonumber\\
T^{x}_{x}         &=&\frac{e^{2}F_{0}^{4}}{2}\left(u f'^{2}-v\right)\nonumber\\
T^{\theta}_{\theta} &=&\frac{e^{2}F_{0}^{4}}{2C^{2}(x)}\left(-f'^{2}+
\frac{\sin^{4}f}{C^{2}(x)x^{4}}\right)\nonumber\\
T^{\phi}_{\phi}&=&{\cal{T^{\theta}_{\theta}}}.
\end{eqnarray}
Using this energy-momentum tensor and our ansatz for the metric,
(\ref{eq:metric}), we obtain Einstein's equations of the form,

\begin{eqnarray}
\label{eq:Einstein}
G^{\mu}_{\nu}   &=& e^2 F_0^2
\delta^{\mu}_{\nu}\bigg[\frac{3B''(x)}{B(x)C^{2}(x)}+\frac{2C''(x)}{C^{3}(x)}+\frac{3B'^{2}(x)}{B^{2}(x)C^{2}(x)}
-\frac{C'^{2}(x)}{C^{4}(x)}+\frac{6B'(x)}{B(x)C^{2}(x)x}+\frac{4C'(x)}{C^{3}(x)x}+\frac{3B'(x)C'(x)}{B(x)C^{3}(x)}\bigg]\nonumber\\
&=& - \frac{\kappa^2 e^{2}F_{0}^{4}}{2}\left(u f'^{2}+v\right)\delta^{\mu}_{\nu}\nonumber\\
G^{x}_{x}&=&e^2 F_0^2 \bigg[\frac{8B'(x)}{B(x)C^{2}(x)x}+\frac{6B'^{2}(x)}{B^{2}(x)C^{2}(x)}
+\frac{2C'(x)}{x C^{3}(x)}+\frac{8B'(x)C'(x)}{B(x)C^{3}(x)}+\frac{C'^{2}(x)}{C^{4}(x)}\bigg]\nonumber\\
&=& \frac{\kappa^2 e^{2}F_{0}^{4}}{2}\left(u f'^{2}-v\right)\nonumber\\
G^{\theta}_{\theta}&=& e^2 F_0^2
\bigg[\frac{4B''(x)}{B(x)C^{2}(x)}+\frac{C''(x)}{C^{3}(x)}+\frac{6B'^{2}(x)}{B^{2}(x)C^{2}(x)}
-\frac{C'^{2}(x)}{C^{4}(x)}+\frac{4B'(x)}{B(x)C^{2}(x)x}+\frac{C'(x)}{C^{3}(x)x}\bigg]
\nonumber \\
&=&\frac{\kappa^2 e^{2}F_{0}^{4}}{2C^{2}(x)}\left(-f'^{2}+\frac{\sin^{4}f}{C^{2}(x)x^{4}}\right)\nonumber\\
\end{eqnarray}
Eqs. (\ref{eq:Einstein}) and (\ref{eq:f}) constitute the equations of
motion for Einstein-Skyrme model consistent with the restrictions
imposed by our ansatz.

\section{Numerical Results}

We want to find solutions for solitonic objects characterized
by a topological charge Q that can be written in terms of an 
integral over the extra-dimensional space as, 
\begin{eqnarray}
	Q= \frac{\epsilon^{ijk}}{24\pi^{2}}\int {\rm tr}\,
	(L_{i}L_{j}L_{k}) \, C^{3}(r) r^{2}drd\Omega_{2}
	= -\frac{2}{\pi}\int \sin^{2}f df
	=-\frac{2}{\pi}\left[\frac{f}{2}-\frac{\sin 2f}{4}
	\right]_{f(0)}^{f(\infty)}\,. \label{}
\end{eqnarray}
One can see that fixing the charge specifies the
boundary conditions for our function $f(r)$. In the following, we will
consider the single charged soliton solution ($Q=1$), which means
that we will take the following boundary conditions for the scalar
field function $f(r)$,
\begin{equation}
f(0) = \pi ~~~~~~~ f(\infty) = 0.
\end{equation}

We want to integrate our equations of motion starting from the core of the
defect, so we still need to specify the conditions for the metric
functions at $x=0$. We demand that our initial data at the origin does not have
any singularity which in turn means that the most general expansion
for the metric functions at $x=0$ should be of the form,
\begin{eqnarray}
B(x)&=&B_{0}+B_{2}x^{2}+O(x^{4})\nonumber\\
C(x)&=&C_{0}+C_{2}x^{2}+O(x^{4})\nonumber\\
f(x)&=&\pi+f_{1}x+f_{3}x^{3}+O(x^{4}).
\end{eqnarray}
Using this expansion in Einstein's equations, (\ref{eq:Einstein}), one
can see that the higher order coefficients can be obtained in terms of
the parameters $f_1,B_0,C_0,\hat \kappa^2 \equiv \kappa^2 F_0^2$ and are given at the lowest
order by, 
\begin{eqnarray*}
B_{2}&=&\frac{B_{0}f_{1}^{4}\hat{\kappa}^{2}}{10C_{0}^{2}}\\
\\
C_{2}&=& -\frac{f_{1}^{2}(5C_{0}^{2}+11f_{1}^{2}){\hat \kappa}^{2}}{40C_{0}}\\
\\
f_{3}&=&-\frac{f_{1}^{3}(162f_{1}^{4}\hat{\kappa}^{2}-5C_{0}^{4}(-16+3\hat{\kappa}^{2})+5C_{0}^{2}f_{1}^{2}(8
    +9\hat{\kappa}^{2}))}{600(C_{0}^{4}+2C_{0}^{4}f_{1}^{2})}\\
\end{eqnarray*}

We numerically solve the system of equations (\ref{eq:Einstein}) and (\ref{eq:f})
using the shooting method, i.e. adjusting $f_{1}$, $B_{0}$ and $C_0$
such that the asymptotical solutions satisfy $f(\infty)=0$ and
$B(\infty)=C(\infty)=1$. We show in Figs. 1-3, a sample of the
numerical solutions found by this procedure.\footnote{Note that we
integrate the solutions to a much longer range in $x$ where we clearly
see the convergence of the different functions to their asymptotic
values. We only plot a limited range in order to show the 
smooth structure of the soliton in the core region.}

We found that similarly to what happens to the Einstein-Skyrme model
in $4D$ \cite{DHS,BC} there is a critical value of $\hat \kappa$,
$\hat \kappa_{c}$, beyond which no more regular solutions exist. Our
numerical investigation reveals that this critical
value is around the order of $\hat \kappa^{2}_{c}\sim\frac{1}{20}$.
\begin{figure}
\label{B-solution}
\centering\leavevmode
\epsfysize=7cm \epsfbox{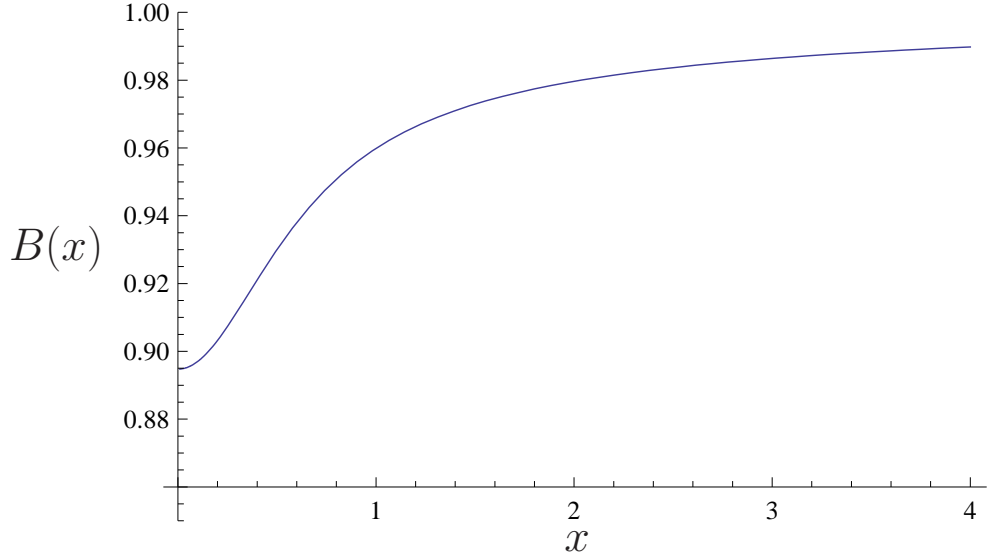}
\put(-160,-10){\Large {\bf {$x$}}}
\put(-370,100){\Large {\bf {$B(x)$}}}
\caption[Fig 1]{Typical form of the $B(x)$ function for the smooth
  3-brane solutions.}
\end{figure}

\begin{figure}
\label{C-solution}
\centering\leavevmode
\epsfysize=7cm \epsfbox{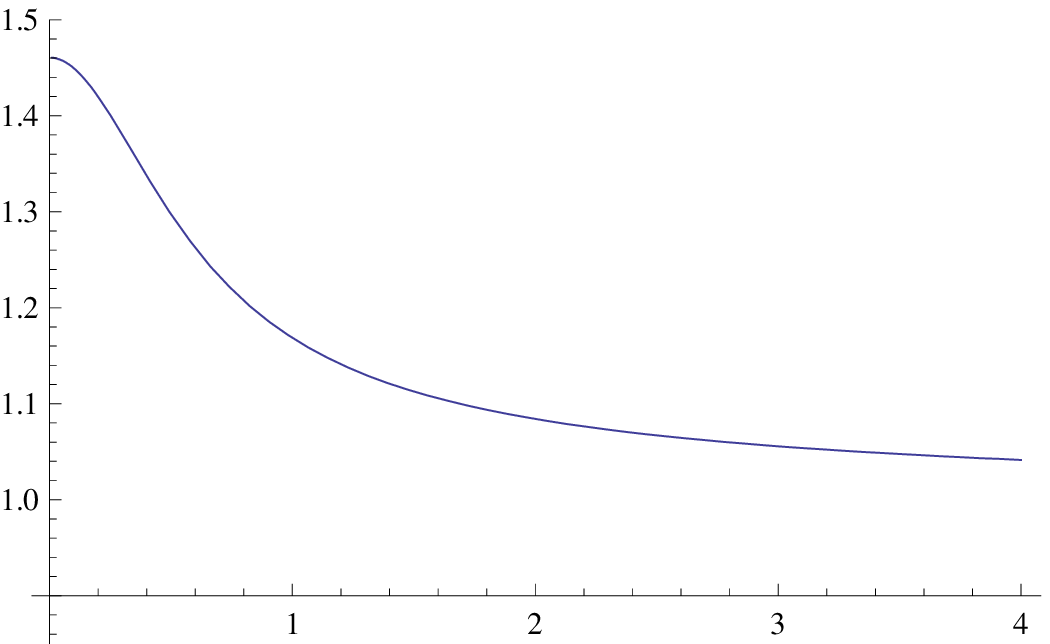}
\put(-160,-10){\Large {\bf {$x$}}}
\put(-365,110){\Large {\bf {$C(x)$}}}
\caption[Fig 1]{Typical form of the $C(x)$ function for the smooth
  3-brane solutions.}
\end{figure}

\begin{figure}
\label{f-solution}
\centering\leavevmode
\epsfysize=7cm \epsfbox{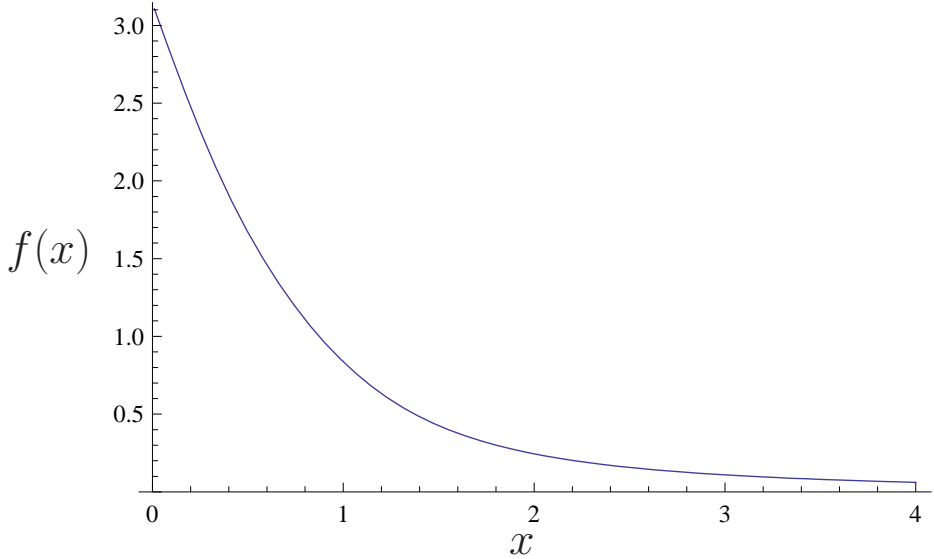}
\put(-160,-10){\Large {\bf {$x$}}}
\put(-350,100){\Large {\bf {$f(x)$}}}
\caption[Fig 1]{Typical form of the $f(x)$ function for the smooth
  3-brane solutions.}
\end{figure}

Another interesting result is that for each value of
$\hat \kappa < \hat \kappa_{c}$ we have two branches of solutions.
At $\hat \kappa=\hat \kappa_{c}$, the two branches merge (See Fig. 4)
such that beyond this point we always find a singularity at some
value of $x$. This is another feature that
is shared by the $4D$ system \cite{BC}.

\begin{figure}
\label{branches}
\centering\leavevmode
\epsfysize=7cm \epsfbox{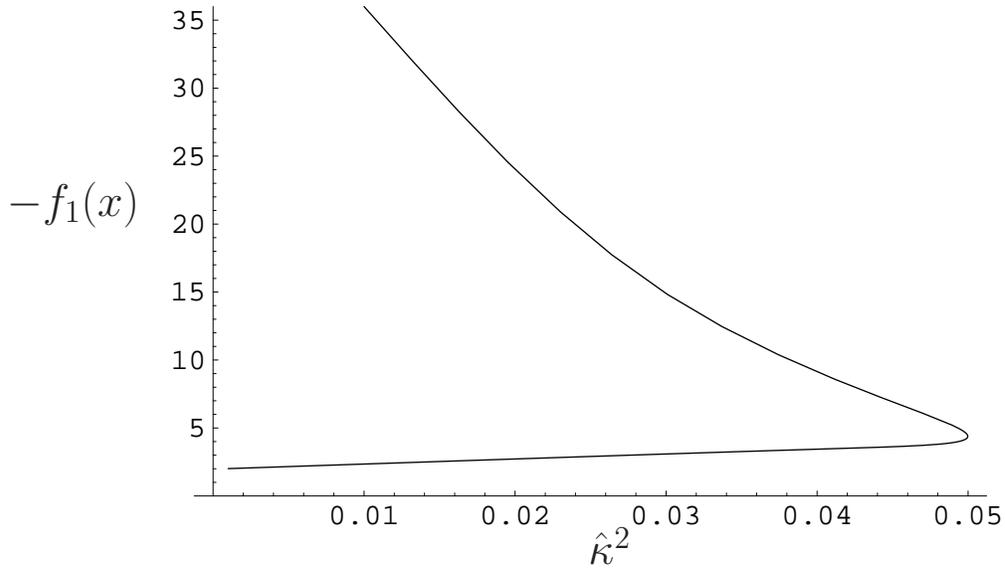}
\put(-160,-12){\Large {\bf {$\hat \kappa^2$}}}
\put(-380,120){\Large {\bf {$-f_1(x)$}}}
\caption[Fig 2]{Two fundamental branches of soliton solutions. The
shooting parameter $f_{1}$ is plotted as a function of $\hat \kappa^{2}$.
We see how the two solutions merge at $\hat \kappa = \hat \kappa_c$.}
\end{figure}

The numerical solutions we found asymtotically approach flat space, and it is
therefore possible to identify the form of the energy momentum tensor
that sources these metrics, in a similar way to what one does in $4D$
spacetime \cite{ADM}. Following \cite{Myers:1999psa,Charmousis:2001hg} 
we obtain,

\begin{eqnarray}
	T^{ADM}_{\mu \nu}=\lim_{r \rightarrow \infty} \frac{1}{2\kappa^{2}}\oint {\hat r}^{i}
	[\eta_{\mu \nu}(\partial_{i}h^{\sigma}_{\sigma}+\partial_{i}h^{j}_{j}
	-\partial_{j}h^{j}_{i})-\partial_{i}h_{\mu \nu}]\, r^{2}d\Omega_{2}
	\label{adm-energy}
\end{eqnarray}
where $h_{AB} = g_{AB} - \eta_{AB}$, denotes the deviation
of our metric from flat space, $\hat r^i$ is the radial unit
vector in the transverse 3-dimensional space and $\mu, \nu , 
\sigma= 0,...,3$ and $i,j=4,5,6$. Using the expression of
our ansatz we obtain in our case,

\begin{equation}
\label{ADM-em}
	T^{ADM}_{\mu \nu}=  \frac{8\pi}{\kappa^{2}}~\eta_{\mu \nu} \lim_{r
          \rightarrow \infty} \left[ r^2 (3 B(r) B'(r) + 2 C(r)
          C'(r))\right]= 
        - {\cal T}_{ADM}~\eta_{\mu \nu} \,
\end{equation}

It is then clear from this calculation that we can read off the value
of the tension of the 3-brane source from the asymtotic behaviour
of the metric. We show in Table I, our numerical results for the
tension (${\cal T}_{ADM}$) computed from the asymtotic form of the
numerical functions $B(x)$ and $C(x)$, together with the values of the 
shooting parameters $B_{0}$, $f_{1}$, $C_{0}$, for a range of $\hat
\kappa$ values. The subscript $u$ denotes the upper branch in Fig. 4.

\begin{table}
\centering
\begin{tabular}{|l|c|c|c|c|c|c|c|c|c|c|}
\hline
 $\hat \kappa^{2}$ & $r_{0}$ & ${\cal T}^{ADM}$ & $f_{1}$ & $C_{0}$ & $B_{0}$
& $r_{0u}$ & ${\cal T}^{ADM}_{u}$ & $f_{1u}$ & $C_{0u}$ & $B_{0u}$ \\
\hline

  0.01   & 0.071614 & 71.1456 & -2.21326 & 1.075 & 0.9796 & 0.120868 & 120.077 & -35.9947 & 4.007 & 0.6524 \\
  0.02   & 0.139140 & 69.1149 & -2.47453 & 1.165 & 0.9550 & 0.177366 & 88.1029 & -21.2452 & 3.506 & 0.6802 \\
  0.03   & 0.202338 & 67.0048 & -2.83941 & 1.283 & 0.9299 & 0.224255 & 74.2627 & -14.3882 & 3.070  & 0.7091 \\
  0.04   & 0.260364 & 64.6652 & -3.43789 & 1.460 & 0.8940 & 0.269732 & 66.9919 & -9.92565 & 2.633 & 0.7450 \\
  0.048  & 0.303528 & 62.8214 & -4.65254 & 1.766 & 0.8445 & 0.303671 & 62.8510 & -6.52400 & 2.140 & 0.7952 \\
  0.0485 & 0.306311 & 62.7438 & -4.88309 & 1.818 & 0.8370 & 0.306296 & 62.7407 & -6.18143 & 2.079 & 0.8028 \\
  0.0487 & 0.307273 & 62.6824 & -5.01605 & 1.847 & 0.8329 & 0.310717 & 63.3849 & -6.02929 & 2.055 & 0.8060 \\
  0.0488 & 0.308737 & 62.8519 & -5.10167 & 1.865 & 0.8300 & 0.308765 & 62.8576 & -5.88885 & 2.024 & 0.8095 \\
  0.0489 & 0.308155 & 62.6052 & -5.22771 & 1.892 & 0.8268 & 0.308099 & 62.5938 & -5.74297 & 1.996 & 0.8131  \\
 \hline
 \end{tabular}
 \caption{We show the values of the shooting parameters for different
   values of the gravitational coupling. We include both branches of
   solutions. The values of $r_0$ are given in $eF_0$ units and ${\cal
     T}^{ADM}$ in $F_0/e$ units.}\label{tableprob}
\end{table}

The occurrence of the two branches in $4D$ has been claimed to be
linked to the existence of similar solutions in the Einstein-Yang-Mills
system in $4D$ \cite{BM}. It is likely that a similar situation may
arise in our higher dimensional case.

\section{Cosmic 3-branes in the isotropic gauge}

As we discussed in the introduction, our brane solutions have
the same symmetry as the cosmic 3-brane gravity solutions discussed
by Gregory \cite{RG}. Furthermore our branes are not charged with
respect to any long range field, and therefore their energy momentum
tensor is very well localized. It is then tempting to identify our
solutions with the smooth out version of the cosmic 3-brane examples.
This possibility was in fact already suggested by Gregory in \cite{RG}.
In the following we will prove that this is indeed the case, by showing
that the metric solutions found numerically in the previous section
match asymptotically the thin wall vacuum solutions found by Gregory.
Let us briefly review the cosmic 3-brane geometries. The solutions
of vacuum Einstein equations found in \cite{RG} relevant for us are of the form,

\begin{equation}
ds^2= F(\hat r)^{2\alpha} \eta_{\mu \nu} dy^{\mu} dy^{\nu} + F(\hat
r)^{2\beta} d\hat r^2 + \hat r^2 F(\hat r)^{2\gamma} d\Omega^2_2
\end{equation}
where
\begin{equation}
F(\hat r) = 1- \frac{\hat r_0}{\hat r}
\end{equation}
and $\alpha = {1\over{2\sqrt{10}}}$, $\beta = -{2\over {\sqrt{10}}}$
and $\gamma = {{1+2\beta}\over 2}$.

These analytic solutions are written in a different gauge
from the numerical ones found in the previous section, but it is
always possible to transform them into the {\it isotropic} gauge of
the form,

\begin{equation}
ds^2= B(r)^2 \eta_{\mu \nu} dy^{\mu} dy^{\nu} + C(r)^2 (dr^2 + r^2 d\Omega_2)~.
\end{equation}
In this gauge, the 3-brane vacuum solutions become,

\begin{equation}
\label{3-brane-isotropic}
ds^2= \left({{4 r - r_0}\over {4 r + r_0}}\right)^{{2\over{\sqrt{10}}}}
\eta_{\mu \nu} dy^{\mu} dy^{\nu} +
\left({{4 r - r_0}\over {4 r}}\right)^4
\left({{4 r + r_0}\over {4 r - r_0}}\right)^{2 + {8\over{\sqrt{10}}}}
(dr^2 + r^2 d\Omega_2)~.
\end{equation}
Using the asymtotic form of this metric and the expression for the
ADM energy-momentum tensor found in (\ref{ADM-em}) we arrive at,

\begin{eqnarray}
	T^{ADM}_{\mu \nu} = -\frac{\sqrt{10}\pi}{\kappa^{2}}r_0\, \eta_{\mu \nu} \, ,
	\label{adm-analytic}
\end{eqnarray}
which depends on the single parameter $r_0$ that completely
characterizes the 3-brane metric solution. We give in Table 1
the results obtained for this parameter in our numerical examples.
Note that our numerical solutions are smooth everywhere 
contrary to the analytic expressions in Eq.(\ref{3-brane-isotropic}) 
that have naked singularities located at
$r=r_0$ and at $r=0$. On the other hand, we will show that the
analytic solutions are in fact a good approximation to our
numerical results in the asymptotic region where $r >> r_0$
but start deviating from them as $r \sim r_0$. In order 
to see this, we proceed in the following way.

\begin{figure}
\centering\leavevmode
\epsfysize=7cm \epsfbox{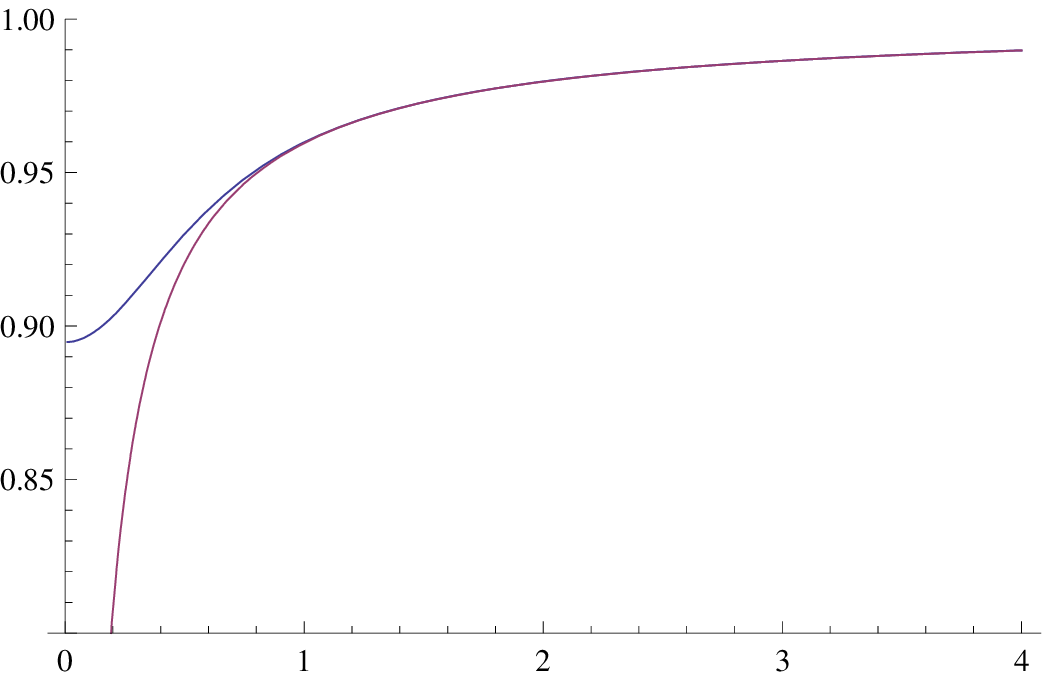}
\put(-150,-10){\Large {\bf {$x$}}}
\put(-340,110){\Large {\bf {$B(x)$}}}
\caption[Fig 1]{Comparison between the numerical $B(x)$ and the
thin wall solution (\ref{3-brane-isotropic}).}
\end{figure}

\begin{figure}
\centering\leavevmode
\epsfysize=7cm \epsfbox{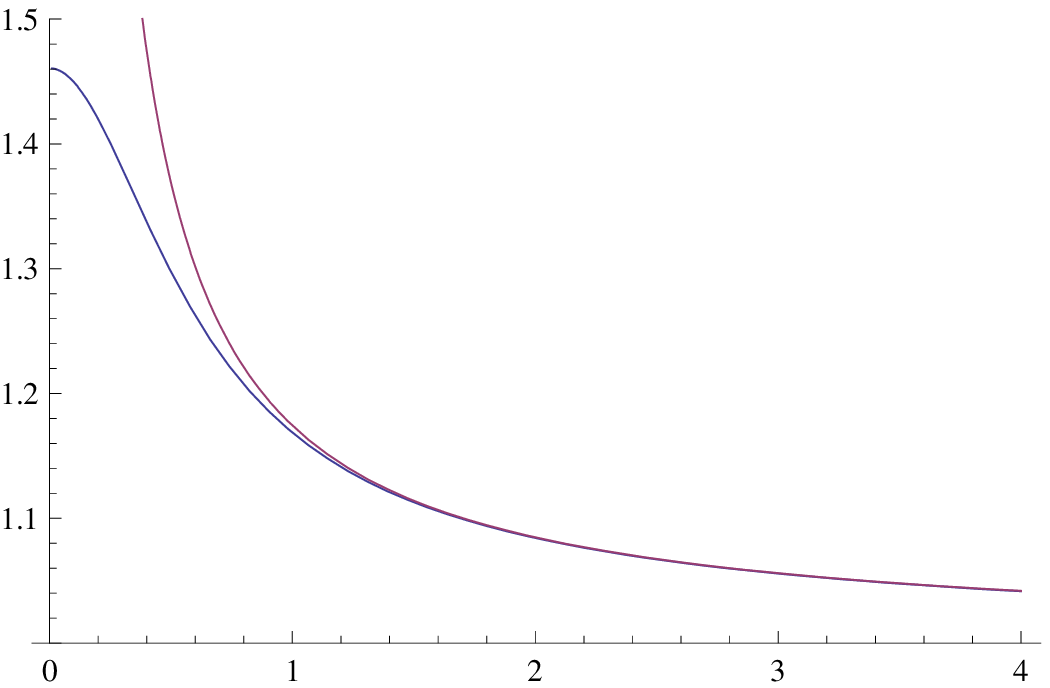}
\put(-160,-10){\Large {\bf {$x$}}}
\put(-340,110){\Large {\bf {$C(x)$}}}
\caption[Fig 2]{Comparison between the numerical $C(x)$ and the
thin wall solution (\ref{3-brane-isotropic}).}
\end{figure}

We first identify, using  Eq.~(\ref{ADM-em}) and
Eq.~(\ref{adm-analytic}), the value of the parameter $r_0$ by looking at 
the asymtotic form of the ADM stress tensor of our numerical solution.
Once this parameter is fixed, it singles out a particular member within
the family of solutions given by Eq.(\ref{3-brane-isotropic}).
Using this parameter we now plot in Figs. 5 and 6 the function
$B(x)$ and $C(x)$ and compare them with the same functions in the
numerical calculations. As we see the asymptotic form of the two
functions, the numerical ones and the analytic vacuum solutions, are
in perfect agreement so we conclude that our numerical solutions do, in fact, 
match their thin wall counterparts.

\section{Conclusions}
We have constructed, using numerical techniques, static 3-brane solitonic solutions in
the $7D$ Skyrme model coupled to gravity. We have shown that their
asymptotic form can be well approximated by the analytic
vacuum solutions of pure Einstein theory obtained in \cite{RG}.
The presence of the energy momentum tensor that makes up the core of
the brane is able to smooth out the singularities that show up in the analytic case,
rendering these solutions completely smooth.

The Skyrme branes obtained in this paper are asymptotically flat and
therefore represent a good candidate for regular braneworlds in the DGP model
in $7D$ \cite{Dvali:2000hr,Dvali:2002pe}. It is also clear that one could
generalize our Skyrme model to higher dimensions in order to
accomodate the DGP braneworld models of higher codimension within
a similar framework to the one studied here. On the other hand,
the properties of the gravitational sector of the braneworld
may be affected by the details of the brane core
\cite{Charmousis:2001hg} so it would be interesting to test these
ideas within our model.

Furthermore, we have found that there is a maximum value of the coupling constant
beyond which smooth solutions are not possible anymore. It is not
clear what type of objects one should obtain
for larger values of the coupling constant. One possibility is to
relax the staticity of the metric. This seems to suggest that the branes 
would start to inflate in a similar way to the solutions found in
\cite{Cho-1,RG-2,Cho-2}. This is certainly a possibility although we note that 
the situation is slightly different from the usual defect solutions 
since we do not have any potential energy in our model, so the 
possibility of having topological inflation \cite{TI} at the core 
does not seem very likely.

We have also found that there are two branches 
of solutions. We expect, as it happens in the
$4D$ case, that the lower (upper) branch corresponds to a stable 
(unstable) configuration, although the stability calculation has to be
performed taking into account the new possible channels opened due to
the extradimensional nature of the solutions we have.

 We hope to come back to these and other interesting issues for Skyrme
 Branes in a future publication.

\section{Acknowledgements}
We would like to thank Gia Dvali, Gregory Gabadadze, Ken Olum, Oriol
Pujolas, Alex Vilenkin for many useful conversations. J.J.B-P is
supported by the National Science Foundation Grant 06533561.


\begin{thebibliography}{99}


\bibitem{BRANES}


  K.~Akama,
  ``An early proposal of 'brane world',''
  Lect.\ Notes Phys.\  {\bf 176}, 267 (1982).


  V.~A.~Rubakov and M.~E.~Shaposhnikov,
   ``Do We Live Inside A Domain Wall?,''
  Phys.\ Lett.\  B {\bf 125}, 136 (1983).


  L.~Randall and R.~Sundrum,
   ``An alternative to compactification,''
  Phys.\ Rev.\ Lett.\  {\bf 83}, 4690 (1999).


\bibitem{domainwalls}
  O.~DeWolfe, D.~Z.~Freedman, S.~S.~Gubser and A.~Karch,
  ``Modeling the fifth dimension with scalars and gravity,''
  Phys.\ Rev.\  D {\bf 62}, 046008 (2000)


\bibitem{strings}
  R.~Gregory,
   ``Nonsingular global string compactifications,''
  Phys.\ Rev.\ Lett.\  {\bf 84}, 2564 (2000).


\bibitem{monopoles}
  E.~Roessl and M.~Shaposhnikov,
   ``Localizing gravity on a 't Hooft-Polyakov monopole in seven dimensions,''
  Phys.\ Rev.\  D {\bf 66}, 084008 (2002).

\bibitem{generalcodimension}
  I.~Olasagasti and A.~Vilenkin,
  ``Gravity of higher-dimensional global defects,''
  Phys.\ Rev.\  D {\bf 62}, 044014 (2000).


\bibitem{Skyrme}
  T.~H.~R.~Skyrme,
  ``A Nonlinear field theory,''
  Proc.\ Roy.\ Soc.\ Lond.\  A {\bf 260}, 127 (1961).

\bibitem{k-defects}
  E.~Babichev,
   ``Global topological k-defects,''
  Phys.\ Rev.\  D {\bf 74}, 085004 (2006).

  E.~Babichev,
   ``Gauge k-vortices,''
  Phys.\ Rev.\  D {\bf 77}, 065021 (2008).


  C.~Adam, J.~Sanchez-Guillen and A.~Wereszczynski,
  ``k-defects as compactons,''
  J.\ Phys.\ A  {\bf 40}, 13625 (2007).

  C.~Adam, N.~Grandi, J.~Sanchez-Guillen and A.~Wereszczynski,
  ``K fields, compactons, and thick branes,''
  J.\ Phys.\ A  {\bf 41}, 212004 (2008).

  C.~Adam, N.~Grandi, P.~Klimas, J.~Sanchez-Guillen and A.~Wereszczynski,
  ``Compact self-gravitating solutions of quartic (K) fields in brane
  cosmology,'', arXiv:0805.3278 [hep-th].

\bibitem{Derrick}
  G.~H.~Derrick,
  ``Comments on nonlinear wave equations as models for elementary particles,''
  J.\ Math.\ Phys.\  {\bf 5}, 1252 (1964).

\bibitem{RG}
  R.~Gregory,
  ``Cosmic p-Branes,''
  Nucl.\ Phys.\  B {\bf 467}, 159 (1996)

\bibitem{DHS} 
  S.~Droz, M.~Heusler and N.~Straumann,
  ``New black hole solutions with hair,''
  Phys.\ Lett.\  B {\bf 268}, 371 (1991).

\bibitem{BC}
   P.~Bizon and T.~Chmaj,
  ``Gravitating skyrmions,''
  Phys.\ Lett.\  B {\bf 297}, 55 (1992).

\bibitem{ADM}
  R.~L.~Arnowitt, S.~Deser and C.~W.~Misner,
  ``The dynamics of general relativity,''
  arXiv:gr-qc/0405109.


\bibitem{Myers:1999psa}
  R.~C.~Myers,
  ``Stress tensors and Casimir energies in the AdS/CFT correspondence,''
  Phys.\ Rev.\  D {\bf 60}, 046002 (1999).

\bibitem{Charmousis:2001hg}
  C.~Charmousis, R.~Emparan and R.~Gregory,
  ``Self-gravity of brane worlds: A new hierarchy twist,''
  JHEP {\bf 0105}, 026 (2001)
  [arXiv:hep-th/0101198].


 \bibitem{BM}
   R.~Bartnik and J.~Mckinnon,
  ``Particle - Like Solutions of the Einstein Yang-Mills Equations,''
  Phys.\ Rev.\ Lett.\  {\bf 61}, 141 (1988).

\bibitem{Dvali:2000hr}
  G.~R.~Dvali, G.~Gabadadze and M.~Porrati,
  ``4D gravity on a brane in 5D Minkowski space,''
  Phys.\ Lett.\  B {\bf 485}, 208 (2000).


\bibitem{Dvali:2002pe}
  G.~Dvali, G.~Gabadadze and M.~Shifman,
  ``Diluting cosmological constant in infinite volume extra dimensions,''
  Phys.\ Rev.\  D {\bf 67}, 044020 (2003).



\bibitem{Cho-1}
  I.~Cho and A.~Vilenkin,
  ``Gravity of superheavy higher-dimensional global defects,''
  Phys.\ Rev.\  D {\bf 68}, 025013 (2003).

\bibitem{RG-2}
  R.~Gregory,
  ``Inflating p-branes,''
  JHEP {\bf 0306}, 041 (2003)



\bibitem{Cho-2}
  I.~Cho and A.~Vilenkin,
  ``Inflating magnetically charged braneworlds,''
  Phys.\ Rev.\  D {\bf 69}, 045005 (2004).



\bibitem{TI}
  A.~Vilenkin,
  ``Topological inflation,''
  Phys.\ Rev.\ Lett.\  {\bf 72}, 3137 (1994);

  A.~D.~Linde,
  ``Monopoles as big as a universe,''
  Phys.\ Lett.\  B {\bf 327}, 208 (1994).
  
  

\end{thebibliography}
\end{document}